\documentclass[journal=jacsat,manuscript=article]{achemso}
\usepackage[version=3]{mhchem}  
\usepackage{graphicx}
\graphicspath{{images/}}
\usepackage{caption}
\usepackage{subcaption}
\usepackage{siunitx}
\usepackage{gensymb}
\usepackage[export]{adjustbox}
\usepackage{subcaption}
\usepackage{wrapfig}
\usepackage{sidecap}
\usepackage[export]{adjustbox}
\usepackage{amssymb}
\usepackage{amsmath}

\usepackage{ulem}
\usepackage[markup=changes]{changes}   

\author{Said Pashayev}
\affiliation[]{Laboratoire Charles Coulomb (L2C), Univ Montpellier, CNRS, Montpellier, France}
\alsoaffiliation[]{Institute of Physics of the Ministry of Science and Education of the Republic of Azerbaijan, Baku, Azerbaijan}

\author{Romain Lhermerout}
\affiliation[]{Laboratoire Charles Coulomb (L2C), Univ Montpellier, CNRS, Montpellier, France}
\alsoaffiliation[]{Univ. Grenoble Alpes, CNRS, LIPhy, 38000 Grenoble, France} 

\author{Christophe Roblin}
\affiliation[]{Laboratoire Charles Coulomb (L2C), Univ Montpellier, CNRS, Montpellier, France}

\author{Eric Alibert}
\affiliation[]{Laboratoire Charles Coulomb (L2C), Univ Montpellier, CNRS, Montpellier, France}

\author{Remi Jelinek}
\affiliation[]{Laboratoire Charles Coulomb (L2C), Univ Montpellier, CNRS, Montpellier, France}

\author{Nicolas Izard}
\affiliation[]{Laboratoire Charles Coulomb (L2C), Univ Montpellier, CNRS, Montpellier, France}

\author{Rasim Jabbarov}
\affiliation[]{Institute of Physics of the Ministry of Science and Education of the Republic of Azerbaijan, Baku, Azerbaijan}

\author{Francois Henn}
\affiliation[]{Laboratoire Charles Coulomb (L2C), Univ Montpellier, CNRS, Montpellier, France}

\author{Adrien Noury}
\affiliation[]{Laboratoire Charles Coulomb (L2C), Univ Montpellier, CNRS, Montpellier, France}
\email{adrien.noury@umontpellier.fr}

\title[An \textsf{achemso} demo]
{Differentiating Confined from Adsorbed Water in Single-Walled Carbon Nanotubes via Electronic Transport}

\begin{document}

\begin{abstract}
In this article, we show that it is possible to differentiate between water adsorbed on the outside of a single-walled carbon nanotube and that confined inside. To this aim, we measured the electronic transport of a carbon nanotube based field effect transistor (CNTFET) constructed with an isolated single carbon nanotube subjected to controlled environments. More precisely, this distinction is made possible by observing the evolution of the transfer characteristic as a function of the electric field imposed by the gate voltage. It appears that the presence of water results in a displacement of the electrical neutrality point, corresponding to a charge transfer between the nanotube and its environment. Using this approach, we demonstrate the existence of 3 types of water molecules: (i) chemically adsorbed on the SiO\textsubscript{2} surface of the substrate, i.e., forming silanol groups; (ii) physically adsorbed outside next to the nanotube; and (iii) confined inside the nanotube. The first one can only be eliminated by high temperature treatment under vacuum, the second one desorbs in a moderate vacuum at room temperature, while the confined water can be removed at room temperature at higher vacuum, i.e. $10^{-3}$ mbar. We also observe that both water adsorption outside and water confinement inside the nanotube are spontaneous and rather fast, i.e. less than 1 minute in our experimental conditions, while removing the water adsorbed outside and confined inside takes much longer, i.e. 40-60 minutes, thus indicating that water confinement is thermodynamically favorable. It is also shown that the metallicity of the nanotube has no qualitative influence on its interaction with water. Our results experimentally prove the stronger affinity of water for the inner surface of CNT than for the outer one.
\end{abstract}

\section{Introduction}
Counter-intuitively water spontaneously adsorbs inside a hydrophobic carbon nanotube (CNT) channel. It was explained by the surface tension of the water being significantly lower than the threshold wetting value of CNT\cite{Dujardin1994_wetting}, and water exhibiting a lower chemical potential inside the single-walled carbon nanotube (SWCNT) than in the bulk\cite{hummer2001_chemical}.\\
Water confinement in nanoscale channels has been found to result in modified water molecule orientation\cite{Pascale_2011} and decreased dielectric permittivity\cite{Fumagali_2018_dielectric}. However, the impact of confined water on the electronic properties of CNT has not been fully understood. It was proposed that confined water might modify the internal electric field, leading to charge polarization of the SWCNT and hence modifying its density of states (DOS)\cite{Cao_2011_internalwetting}.\\ 
To add complexity, two types of water states can exist when studying water adsorption on individual SWCNT deposited on SiO\textsubscript{2} substrate: (i) chemisorbed water, which forms Silanol groups at the SiO\textsubscript{2} surface and hence at the interface with CNT and (ii) physisorbed water, which is weakly adsorbed water molecules either directly onto the CNT surface or bonded with Silanol groups. It was reported that physisorbed water can be easily removed by pumping under a vacuum at room temperature for a short period of time, while removing chemisorbed water requires more than 200 °C and secondary vacuum\cite{Kim_2003_main}.\\
To the best of our knowledge, a clear discrimination of the impact of water confined inside or adsorbed outside is missing\cite{Agrawal_2017}. However such a discrimination is not made possible on macroscopic samples since one may expect the CNT response to water adsorption to be extremely broad when working with a distribution of different diameters. Therefore, it is necessary to investigate the impact of water at the level of an individual CNT.\\
It is well known that the intrinsic electronic properties (carrier mobility, neutrality points) of a SWCNT as well as the extrinsic properties (Schottky barrier between SWCNT and metal) are very sensitive to the environment\cite{Heller_2009_environment,Cui_2003_contact}. For instance, exposure of the CNTFET to ambient atmosphere resulted in the appearance of a hysteretic behavior\cite{Ong_2011_hyst,Sharf_2012_hyst}. In another case where the CNTFET was exposed to various salts, it was also reported that the CNTFET conductance was modified. Four possible mechanisms were proposed to account for the modification of the CNT transfer characteristic upon analyte adsorption: electrostatic gating or doping, Schottky barrier modulation, capacitance change, and charge mobility variation\cite{Heller2008}. Electrostatic gating manifests as a shift of the conductance along the gate voltage axis because of the CNT charge doping. Adsorbed molecules can modulate the local work function at the metallic electrode and CNT interface, tuning the Schottky barrier and resulting in an asymmetric change for hole and electron branches in the CNT transfer characteristic. The capacitance effect occurs when the gate capacitance changes due to the permittivity of the adsorbed molecules, resulting in a change in the slope of both branches. Finally, the adsorbed molecules can change the carrier mobility leading to an increase or a decrease of the conductance of either one or both the hole and electron branches.\\ 
Water-electron coupling has been studied, both with molecular dynamic simulation (MD) and experiments mostly when water was adsorbed on the outer wall of closed individual CNTs or CNTs network. While studying water adsorption on a SWCNT with MD, it was reported that water decreases its electrostatic gating, mobility, and capacitance\cite{Pati_2002_216}. It was assigned to doping or capacitance between water and CNT\cite{Takao_2003_217,Roberts_2009_218}. Experimentally it was reported that water adsorption on CNTs alters its threshold voltage and ON-OFF ratio\cite{Lin_2006_219}. The electron-donating nature of water was reported when a network of closed SWCNT was exposed to water\cite{Mudimela_2012_220}. It was also shown that networks of semiconducting SWCNTs were more sensitive than metallic ones, highlighting the dependence of the water-electron coupling to the metallicity of the nanotubes. This outcome was explained because of the lower DOS around the Fermi level in metallic tubes compared to the valence band edge in semiconducting ones\cite{lee2005_221}. To summarize, liquid water adsorption outside the CNTFET has been found to increase hysteresis and shift gate voltage neutrality points towards greater negative voltages, but the underlying mechanism of water-CNT coupling is still debated, between doping or capacitance effect\cite{Pati2002_216,kauffman2008_222,Na_2005_211}.\\
In this work, we investigate the water interaction with a CNTFET made of an individual SWCNT. We measure the impact of water adsorption and confinement by comparing the behavior of the SWCNT-FET when the tube is first closed and then opened allowing to clearly distinguish between the different water states. Interaction is observed as a shift of the gate voltage neutrality point in the CNT transfer characteristics, that we found to be dominated by doping rather than capacitance effect. While studying water adsorption and desorption, we distinguish three water adsorption states: 1) water physisorbed outside the CNT, 2) water confined inside the CNT, and 3) water chemisorbed (i.e. Silanol groups) at the SiO\textsubscript{2} dielectric surface at close proximity to the CNT. We observe that water adsorption outside and confinement inside is spontaneous and fast, i.e. less than 1 minute in our experimental condition, while water desorption takes 40-60 minutes and requires vacuum and/or annealing depending on the adsorption site considered. This time difference between adsorption and desorption reflects a large entropic contribution\cite{Pascale_2011}. We also show that the metallicity of the nanotube has no qualitative influence on its interaction with water.

\section{Materials and Methods}
\begin{figure}[htbp]
\centering
  \includegraphics[width=1\textwidth]{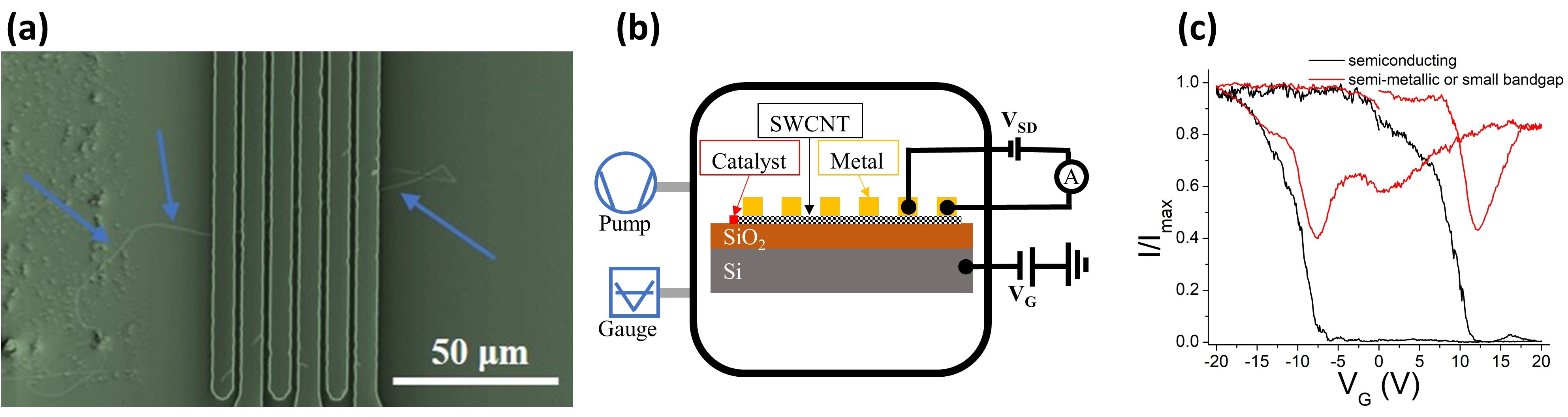}
  \captionsetup{font={small, it}}
  \caption{Device fabrication and characterization: (a) SEM image of a CNTFET. The nanotube (blue arrows) is grown by CVD from the catalyst patterns (visible on the left). Ti and Pt electrodes are deposited on the CNT by E-beam evaporation; (b) Customized chamber to measure the device under a controlled atmosphere; (c) transfer characteristic curves measured for a semiconducting and a semi-metallic or small band gap CNT in air.}
  \label{fig:Figure1}
\end{figure}

Carbon nanotubes are grown by Chemical Vapor Deposition (CVD) on a Si/SiO\textsubscript{2} wafer (doped  Si with SiO\textsubscript{2} layer obtained either by dry oxydation (300 nm) or wet oxydation (2 \textmu m). The nanotubes are synthesized from a Fe-based catalyst solution\cite{JOURDAIN20132}. Scanning Electron Microscope (SEM) imaging allows to locate an individual SWCNT long enough to be part of FET. The selected nanotube is electrically contacted by the deposition of Titanium and Platinum electrodes with 10 nm and 90 nm thicknesses respectively. Figure \ref{fig:Figure1}a presents a SEM image of the final device, showing several metallic electrodes of 5 \textmu m width and spaced by a 5 \textmu m gap, deposited on the SWCNT. A global gate is achieved by contacting the doped Si substrate with silver paste. The transfer characteristics of the SWCNT-FETs are measured in air using a probe station and source measure unit (SMU). In order to characterize the device in a controlled atmosphere or vacuum, a customized chamber has been designed as shown in Figure \ref{fig:Figure1}b.
\\
The transfer characteristic of the CNT is measured by sweeping the gate voltage while applying a constant source-drain bias voltage, i.e. V\textsubscript{SD} = 10 mV. Figure \ref{fig:Figure1}c shows the typical response of a semiconducting (black) and a semi-metallic or small band gap (red) CNT in air. Hysteresis is observed in the CNT transfer characteristic and is related to the charge (electrons or holes) trapping in the environment: on the substrate, at the substrate-CNT interface, and on the CNT\cite{mi13040509}.\\
A reliable and consistent study of the impact of water requires working with individual SWCNTs, while exposing successively their outside and inside to water. Luckily CVD-grown CNTs are ends closed with fullerene-like caps, and it is necessary to intentionally open their ends in order to fill the CNTs. This can be achieved by electrical breakdown. We open CNT sides by applying voltage between selected electrodes aiming to cut CNT at a specific position. Flowing current leads to Joule heating of the CNT and cutting of the CNT, observed as a sudden breakdown in the current-voltage curve (Figure S1). Applying a large current in the range of 10-50 \textmu A leads to heating the CNT up to 1300-1600 K\cite{Khasminskaya_2014_temperature}. In this step, we clearly distinguish between individual SWCNT and other cases, e.g. multiwall or bundles. When the breakdown current is lower than 30 \textmu A and presents only 1 jump on both extremities of the nanotube, the studied tube can be considered as an individual SWCNT (Figure S1). In what follows, we only focus on individual SWCNT.

\section{Coupling mechanism between electrons and water molecules}
\begin{figure}[htbp]
\centering
  \includegraphics[width=1\textwidth]{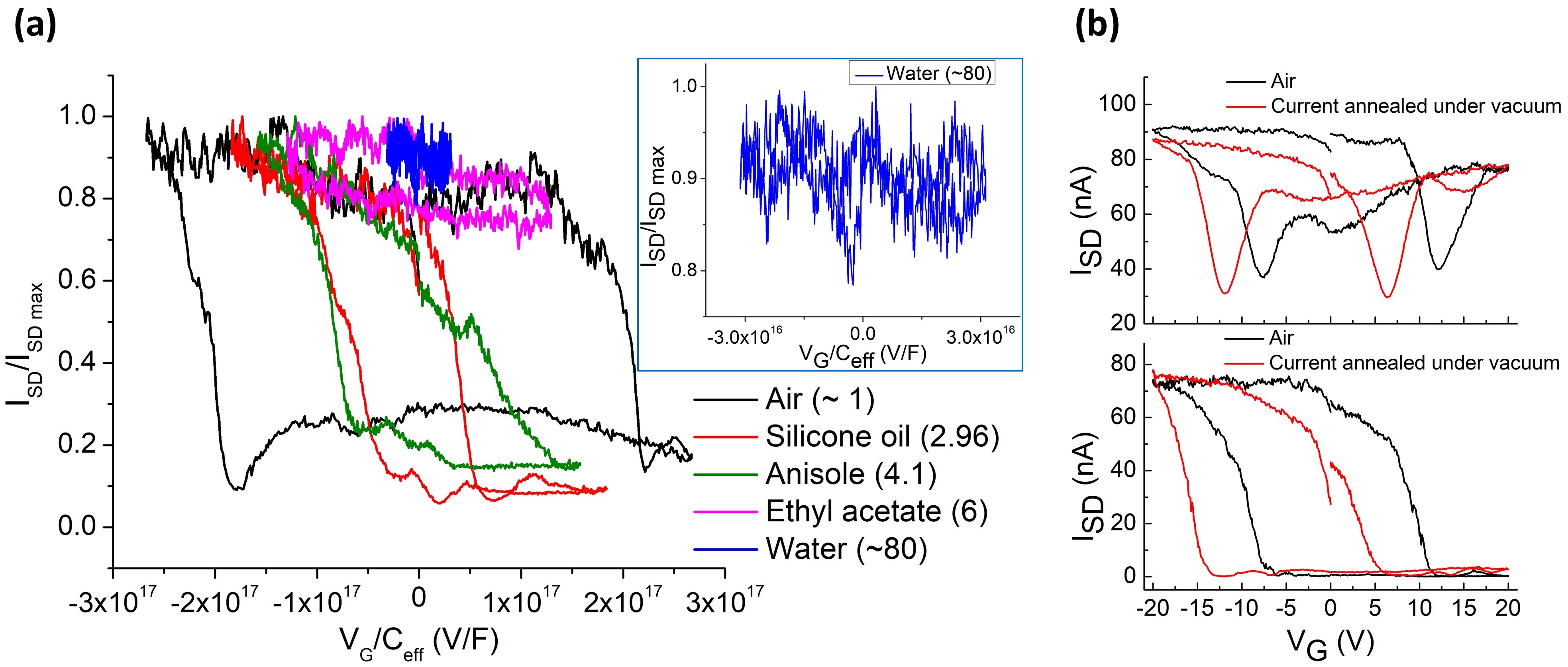}
  \captionsetup{font={small, it}}
  \caption{The transfer characteristic curve of the SWCNT upon exposure to different environments: (a) Closed individual SWCNT exposed to liquids with different dielectric constants. Normalized current versus gate voltage rescaled by the capacitance formed with different liquids. Inset is a zoom at low
$V\textsubscript{G}/C\textsubscript{eff}$ in the case of water; (b) $I\textsubscript{SD}$ as a function of $V\textsubscript{G}$ for two SWCNTs made of either semi-metallic or small bandgap (top) or semiconducting (bottom) individual nanotubes exposed to air (black curves) and after current annealing (T \textgreater 1000 K) under vacuum (red curves).}
  \label{fig:Figure2}
\end{figure}

In this work, we demonstrate that the coupling between electrons and water is dominated by doping rather than capacitive effects.\\
In Figure \ref{fig:Figure2}a, we compare the response of a closed individual SWCNT in air and soaked with milli-Q water. We observe that the ON/OFF transition disappears under water. Among the 4 possible mechanisms reported in literature\cite{Heller2008} , only two can explain such an outcome: (i) the capacitance effect due to the high dielectric constant of water that would rescale the gate axis, therefore effectively shifting the ON/OFF transition or (ii) the doping due to the increased amount of charge trapping in the CNT environment that would shift the neutrality point.\\

In order to verify the first hypothesis, we soaked the closed CNTFET with nonaqueous liquids of increasing dielectric constants. If the conductance change was only due to the dielectric constant of the soaking liquids, then it would be possible to obtain a master curve by rescaling the V$\textsubscript{G}$ axis by the so formed capacitance, since the CNTFET conductance is proportional to the applied electric field, i.e. G$\textsubscript{CNT}$=f(q$\textsubscript{CNT})$ with q$\textsubscript{CNT}$=C$\textsubscript{eff}$V$\textsubscript{G}$. Here C$\textsubscript{eff}$ is the effective gate capacitance, which is a function of both the dielectric constant of the SiO\textsubscript{2} and the dielectric constant of the liquid environment C$\textsubscript{eff}$=g($\epsilon\textsubscript{SiO\textsubscript{2}}$, $\epsilon\textsubscript{r})$, where $\epsilon\textsubscript{SiO\textsubscript{2}}$ is the permittivity of the dielectric layer, $\epsilon\textsubscript{r}$ is the dielectric constant of the soaking liquid. The exact value of C$\textsubscript{eff}$ for each liquid with a different dielectric constant is obtained by numerical simulation. The obtained values are presented in Table S1. Figure \ref{fig:Figure2}a reports the normalized current versus the rescaled gate voltage for the tested liquids. The current normalization is to better track shifts of gate axis. It can clearly be seen that the transfer functions do not overlap. Therefore, the dielectric constant of the liquid is not the dominant parameter to explain the change in the transfer characteristic. Thus, it can be assumed that doping/gating is likely to be the main coupling mechanism between water and the SWCNT-FET.\\
In what follows, we confirm the doping nature of the coupling by comparing the transfer characteristic curve of the CNTFET in air and after annealing under secondary vacuum, that is when the CNT can be considered fully dehydrated. 
Figure \ref{fig:Figure2}b presents the transfer characteristics of a CNTFET made of an individual SWCNT with different metallicities when exposed to air with relative humidity in the range of 40-50\% and after current annealing under $10^{-3}$ mbar vacuum. We observe that the main difference between water-exposed and water-free devices is the shift of the gate voltage neutrality point, independently of the metallicity of the CNT. This observation can be linked to the doping of the CNT by water molecules located outside. Similarly, we also observe that doping dominates upon exposing the closed SWCNT to different humidity levels (Figure S2).

\section{Differentiating water confined inside from that adsorbed outside the individual SWCNT}
\begin{figure}[htbp]
\centering
  \includegraphics[width=1\textwidth]{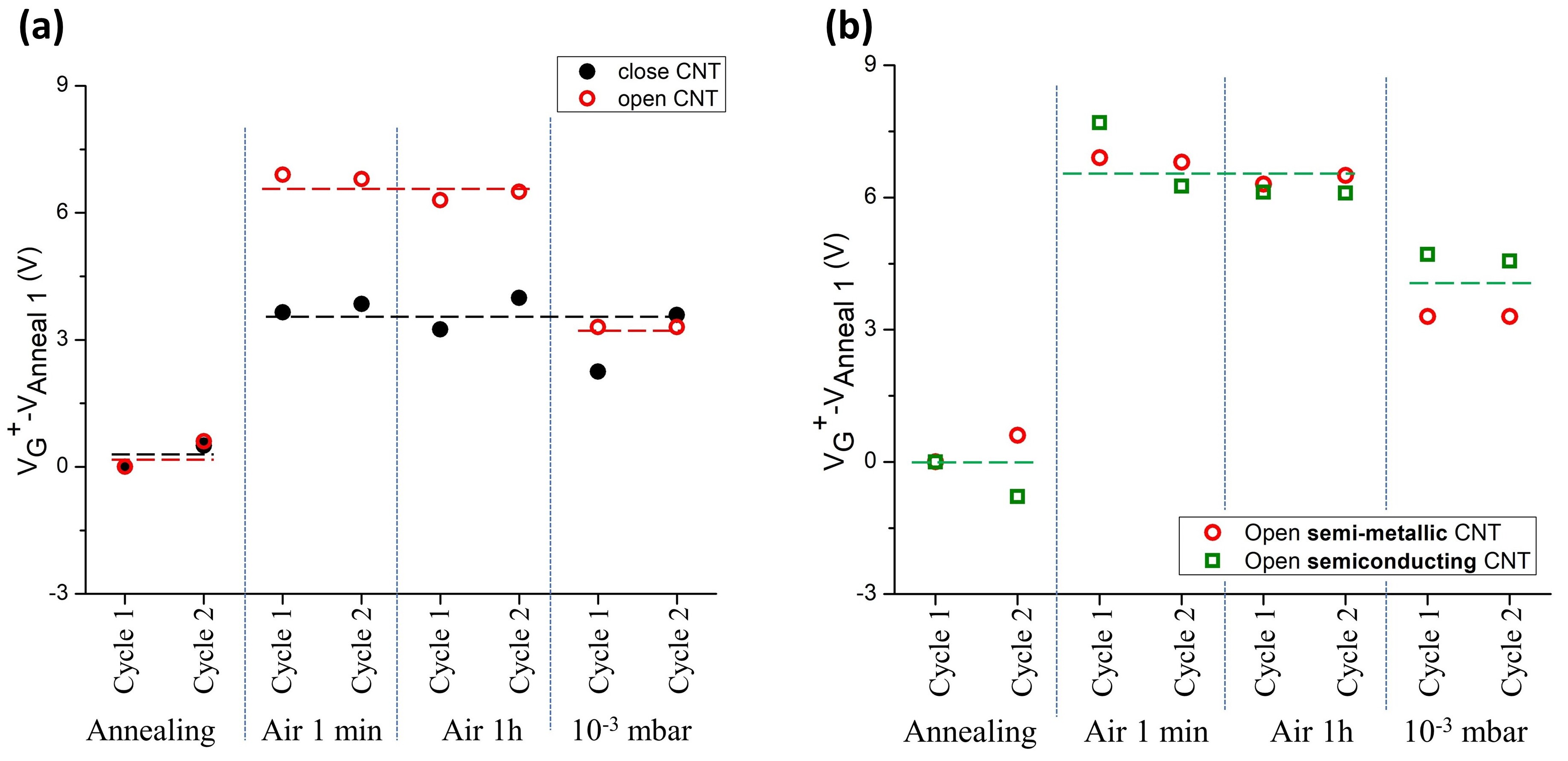}
  \captionsetup{font={small, it}}
  \caption{Differentiating water confined inside from that adsorbed outside an individual SWCNT and the impact of metallicity. Closed and open tubes are represented by full and open symbols respectively. Symbols are experimental data, horizontal and vertical dashed lines are guides to the eye. (a) Individual SWCNT exposed to different environments: after current annealing under $10^{-3}$ mbar vacuum, $\sim$ 1 min exposure to ambient air, 1h exposure to ambient air, and under $10^{-3}$ mbar vacuum at room temperature, (b) comparison of 2 open individual SWCNTs with different metallicities, exposed to the same environments as in (a).}
  \label{fig:Figure3}
\end{figure}

We now demonstrate that it is possible to qualitatively differentiate water adsorbed outside from that confined inside by comparing the CNTFET response of a given device before and after opening of the tube. The only way to measure the impact of water adsorption is to set a reference state for each CNTFET. In our case, this reference state should be that of the CNTFET after annealing under vacuum, i.e. when the device is completely dry and the concentration of silanols present on the substrate surface is reduced to a minimum. Figure \ref{fig:Figure3}a shows the relative change of V$\textsubscript{G}\textsuperscript{+}$ (corresponding to the neutrality point on the positive side, by downward swap) of closed and opened tubes under different successive environments: (i) after current annealing under $10^{-3}$ mbar vacuum, (ii) after about 1 min exposure to ambient air, (iii) after 1h exposure to ambient air, and (iv) at room temperature under $10^{-3}$ mbar. We repeated the exposure cycles several times to ensure reproducibility (Figure S3). The upward (V$\textsubscript{G}\textsuperscript{+}$) and downward (V$\textsubscript{G}\textsuperscript{-}$) neutrality points are physically equivalent, but sometimes we can only track one, due to the limited measurement window.\\
The main difference is observed between closed and open cases after the CNTFET has been placed under a $10^{-3}$ mbar vacuum. While vacuum does not make any observable impact on the closed CNTFET, it significantly changes V$\textsubscript{G}\textsuperscript{+}$ when the nanotube is open. This shows that the change is due to the water confined inside the CNT. The initial increase of V$\textsubscript{G}\textsuperscript{+}$ when the CNTFET is exposed to air, whether the tube is closed or open, can be attributed to doping by the water molecules adsorbed outside, either chemically or physically. We also tested the device when it was exposed to He just after the annealing under vacuum (Figure S4). The absence of change in the transfer characteristic confirms that the previous observed impact is related to air (oxygen and water molecules).\\

Exposure to ambient air for 1 hour did not lead to any change in both open and close cases. This outcome indicates that the impact of water adsorption is spontaneous and rather fast, both outside and inside. In addition, we also tested the opened CNTFET when it is first exposed to liquid water and then subsequently dried at room temperature using a N$\textsubscript{2}$ flow (Figure S5). It is then shown that exposure to liquid water and evaporation has the same impact as exposure to ambient air. This pinpoints the dominant effect of water rather than that of oxygen in the response of our device. It also emphasizes that the removing of the liquid water using a dry N$\textsubscript{2}$ flow is not enough to desorb the water molecules confined inside the tube, highlighting the stable confinement of water inside the nanotube. However, under $10^{-3}$ mbar vacuum, the closed CNTFET exhibits no significant change, while the open CNTFET exhibits a significant shift in the gate voltage neutrality point. This shift can therefore be ascribed to the desorption of the confined water. The difference observed between closed and open cases of the same individual CNTFET clearly shows that confined water inside can be discriminated from that of water adsorbed outside.\\
Finally, both open and closed tubes return to their initial state after annealing under vacuum. Assuming that there is no more physically adsorbed water under $10^{-3}$ mbar vacuum at room temperature, it means that the shift of V$\textsubscript{G}\textsuperscript{+}$ results from the disappearance of the silanol groups at the substrate surface surrounding the tube. When comparing the V$\textsubscript{G}\textsuperscript{+}$ shift induced by the silanol group and that of the water molecules confined inside, it turns out that the impact of both are rather significant and quantitatively similar.\\
We now estimate the amount of charge on the nanotube surface induced by the water molecules. From a simple electrostatic capacitor model, we can write q$\textsubscript{NT}$=C$\textsubscript{eff}$V$\textsubscript{G}$, hypothesizing that the dominant effect of water-nanotube interaction is doping as demonstrated earlier (that is, neglecting capacitance change), and that all the potential difference transfers into a charge (neglecting quantum capacitance). We can therefore write that dq$\textsubscript{NT}$=C$\textsubscript{eff}$dV$\textsubscript{G}$, that relates the charge due to water dq$\textsubscript{NT}$ to the measured neutrality point shift dV$\textsubscript{G}$. To better compare with previous report\cite{Manghi_2021,Grosjean2019} we compute the change per unit area. The surface of the nanotube is S$\textsubscript{NT}=2\pi$R$\textsubscript{NT}$L$\textsubscript{NT}$, where L$\textsubscript{NT}$ is the length of the nanotube between 2 successive electrodes and  R$\textsubscript{NT}$ is the nanotube radius. We assume that R$\textsubscript{NT}$=1 nm and from our design L$\textsubscript{NT}$=5 $\mu$m, therefore we get $\rho\textsubscript{NT}$=0.06 C.m$\textsuperscript{-2}$. While taking into account the number of atoms of a nanotube $\sim 2*10\textsuperscript{19}$ atoms.m$\textsuperscript{-2}$, we can calculate the charge per carbon atom equal to $\sim$ 0.02 e$\textsuperscript{-}$/carbon. The estimated value is in line with the literature\cite{Manghi_2021}. We note that the change is negative. This result is possible with the mechanism of OH$\textsuperscript{-}$ adsorption as proposed by Grosjean et al.\cite{Grosjean2019}\\
In order to verify if the water-electron coupling is dependent on the electron DOS of the nanotube, we select two individual open SWCNTs with different metallicities, one of them being semiconducting the other one semi-metallic. The same protocol is repeated with two cycles (Figure \ref{fig:Figure3}b). We observe the same behavior for both CNTFET, thus showing that the impact of water adsorption on the CNTFET is quantitatively and qualitatively independent of the metallicity of the nanotube.\\
So far, we focused on individual SWCNTs. We also tested the nanotubes which were either bundle or MWCNT, as assigned at the nanotube opening step. When this group of CNTFET are treated in different environments (Figure S6), we do not observe any significant difference between open and closed cases. We attribute this behavior to the dominant electronic response of the outermost wall in the case of DWCNT and MWCNT which is not influenced by the presence of water confined inside the central channel. In addition, in the case of bundle, it is more difficult to distinguish the impact of water because of the diversity of the tubes, different water adsorption sites, and the poor contact with the electrodes. To be clear no trend can be drawn from the measurements when the tube is not an individual SWCNT.\\
To sum up, the impact of the water confined inside the CNTFET can be clearly differentiated from that adsorbed outside, only in the case of individual SWCNTs, and the charge transfer observed is independent of the metallicity of the nanotube.

\section{Water desorption}
\begin{figure}[htbp]
\centering
  \includegraphics[width=1\textwidth]{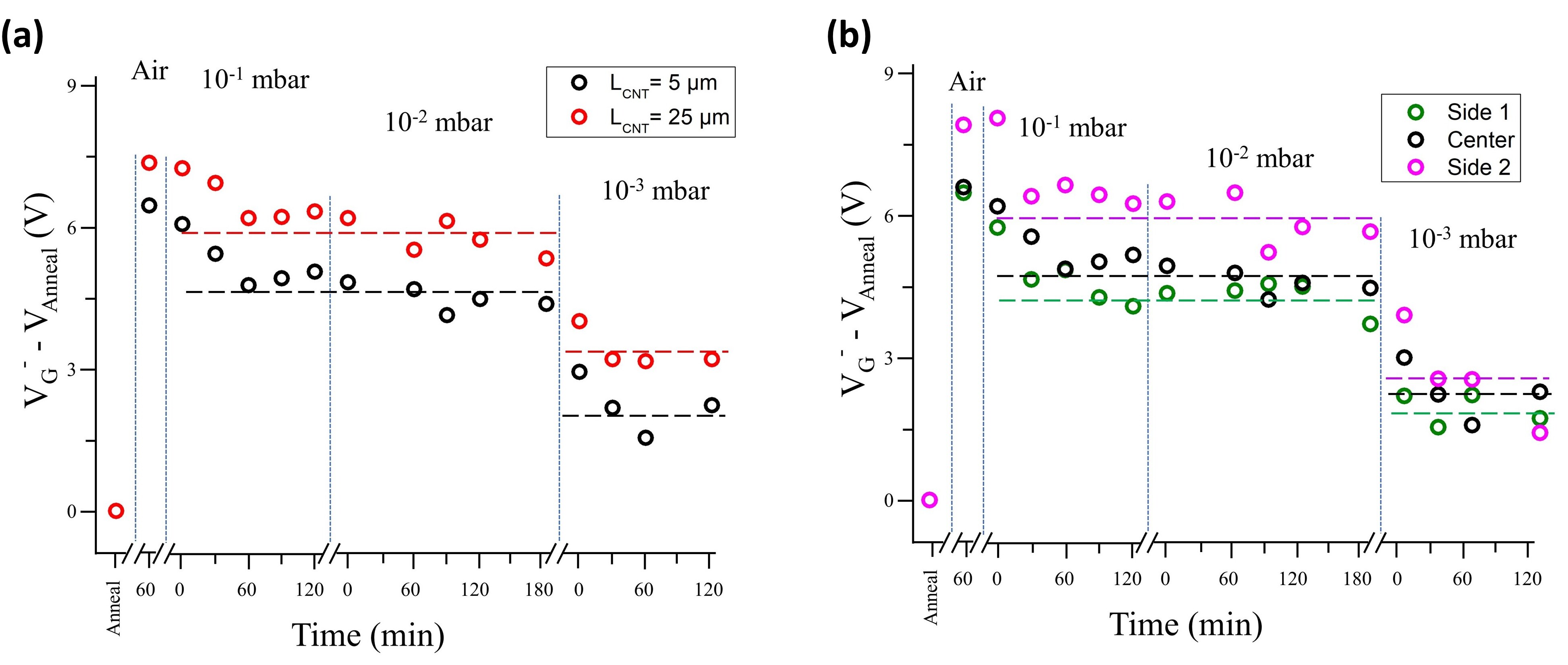}
  \captionsetup{font={small, it}}
  \caption{The change of V$\textsubscript{G}\textsuperscript{-}$ as a function of time for different vacuum pressures. Symbols are experimental data, horizontal and vertical dashed lines are guides to the eye. (a) water desorption from CNTFET with different channel lengths (5 \textmu m and 25 \textmu m) of the same individual SWCNT; (b) water desorption from the same channel length (5 \textmu m) of the same individual SWCNT measured near to the opened extremities and in the center.}
  \label{fig:Figure4}
\end{figure}

Let us now investigate the desorption of water. Figure \ref{fig:Figure4} reports the behavior of the annealed, open CNTFET when exposed to ambient air and then to an increasing level of vacuum, while monitoring the relative change in the neutrality point. Once a given vacuum is stabilized, the transfer characteristic curve is measured every 30 minutes until it remains unchanged. Then, the pressure is decreased to a lower level. We measured water desorption on different nanotube lengths (5-25 \textmu m) of the same individual open SWCNT, and also the response of different 5 \textmu m long sections of the SWCNT.\\ 

When studying the impact of water desorption on sections with different lengths (5 \textmu m and 25 \textmu m) of the same individual opened SWCNT (Figure \ref{fig:Figure4}a), it is observed that all sections behave qualitatively similarly. Whatever their length, we observe a slight decrease of the neutrality point that is not instantaneous when the pressure is $10^{-1}$ mbar. Then at $10^{-2}$ mbar, no significant change is observed. While at $10^{-3}$ mbar there is again another slow decrease of V$\textsubscript{G}$. Thus we can assume that the first step at $10^{-1}$ mbar corresponds to the removal of the water molecules physisorbed at the external surface of the tube and in its vicinity at the substrate surface. The second step at $10^{-3}$ mbar can be assigned to the desorption of water molecules confined inside the tube. The time dependence of V$\textsubscript{G}$ shows that desorption of water physisorbed both outside and inside the tube is not instantaneous. Furthermore, the fact that the 5 \textmu m and 25 \textmu m sections behave the same, implies that the mechanism that governs the desorption does not depend on the length explored; (i) it is straightforward in the case the water physisorbed outside the tube, (ii) in the case of the water confined inside the tube, it means that the desorption is governed by the extraction of the water molecules and not by the reorganization of the water molecule remaining inside the tube. In other words, the main energy barrier is the one associated to the extraction of the water molecules from the SWCNT extremities, rather than the one associated to the diffusion of water molecules inside the tube. This is in line with the fact that water confinement is thermodynamically favorable\cite{Pascale_2011}.\\
The data reported in Figure \ref{fig:Figure4}b for various 5 \textmu m sections is consistent with the results reported in Figure \ref{fig:Figure4}a and the conclusion we have just drawn.\\ 
We did not observe any difference between the sections of the device located close to the opened extremities and the sections located in the center, far from the extremities. It confirms that the main energy barrier for water molecules is linked to desorption at the SWCNT extremities rather than diffusion along the nanotube.

\subsection{Conclusion}
We showed that it is possible to differentiate water confined inside a SWCNT from that adsorbed outside, by monitoring the transfer characteristic of CNTFET based on individual SWCNT. Water/CNT coupling mechanism was assigned to be dominated by doping, i.e. the charge transfer from the water molecules to the CNT, thanks to the shift of the neutrality point of the transfer characteristic. Water adsorption outside and confinement inside is spontaneous and rather fast (i.e. less than 1 min in our experimental condition), while physically adsorbed water is desorbed in 40-60 minutes at room temperature under vacuum. The impact of water adsorption on the CNT is independent of its metallicity. We identify three water adsorption sites that we attribute to (i) physisorbed water, (ii) water confined inside CNT, and (iii) chemisorbed water, each with distinct adsorption energy. In addition, we found that the energy barrier for water molecules desorption is governed by their extraction from the SWCNT extremities rather than by their diffusion inside the tube.\\
Last but not least, our investigation clearly shows that, under our experimental conditions, it was possible to gain reproducible and reliable outcomes if and only if the CNTFET was made with an individual SWCNT.

\subsection{Author Contributions}
A.N, S.P., and F.H. conceived the experiment; S.P. fabricated the samples with the help of Re.J., R.L., E.A., and C.R.; S.P. carried out the experiment with inputs from A.N. and F.H.; N.I. performed the numerical simulation; A.N., F.H., and Ra.J. supervised the work. All authors have read and agreed to the manuscript.

\subsection{Conflicts of interest}
There are no conflicts of interest to declare.

\begin{acknowledgement}
We acknowledge financial support from CNRS-MITI (France) through a 'Momentum' grant. SP acknowledges support from the “State Program 2019-2023” (Azerbaijan) for the Ph.D. scholarship. The work was partly supported by the French Renatech network.
AN acknowledges funding from ANR, project number ANR-22-CE09-0004 and funding from University of Montpellier program 'Soutien à la recherche 2022'. We thank Aloise Degiron for help with Comsol simulations. We thank Saïd Tahir and Vincent Jourdain for help with CNT growth. A CC-BY public copyright license has been applied by the authors to the present document and will be applied to all subsequent versions up to the Author Accepted Manuscript arising from this submission, in accordance with the grant’s open access conditions.
\end{acknowledgement}
\suppinfo{} 
\bibliography{My_biblio}
\end{document}


\tableofcontents
\setcounter{tocdepth}{1}

\section{1. CNT sides opening and selection}
\begin{figure}[htbp]
\centering
  \includegraphics[width=0.5\textwidth]{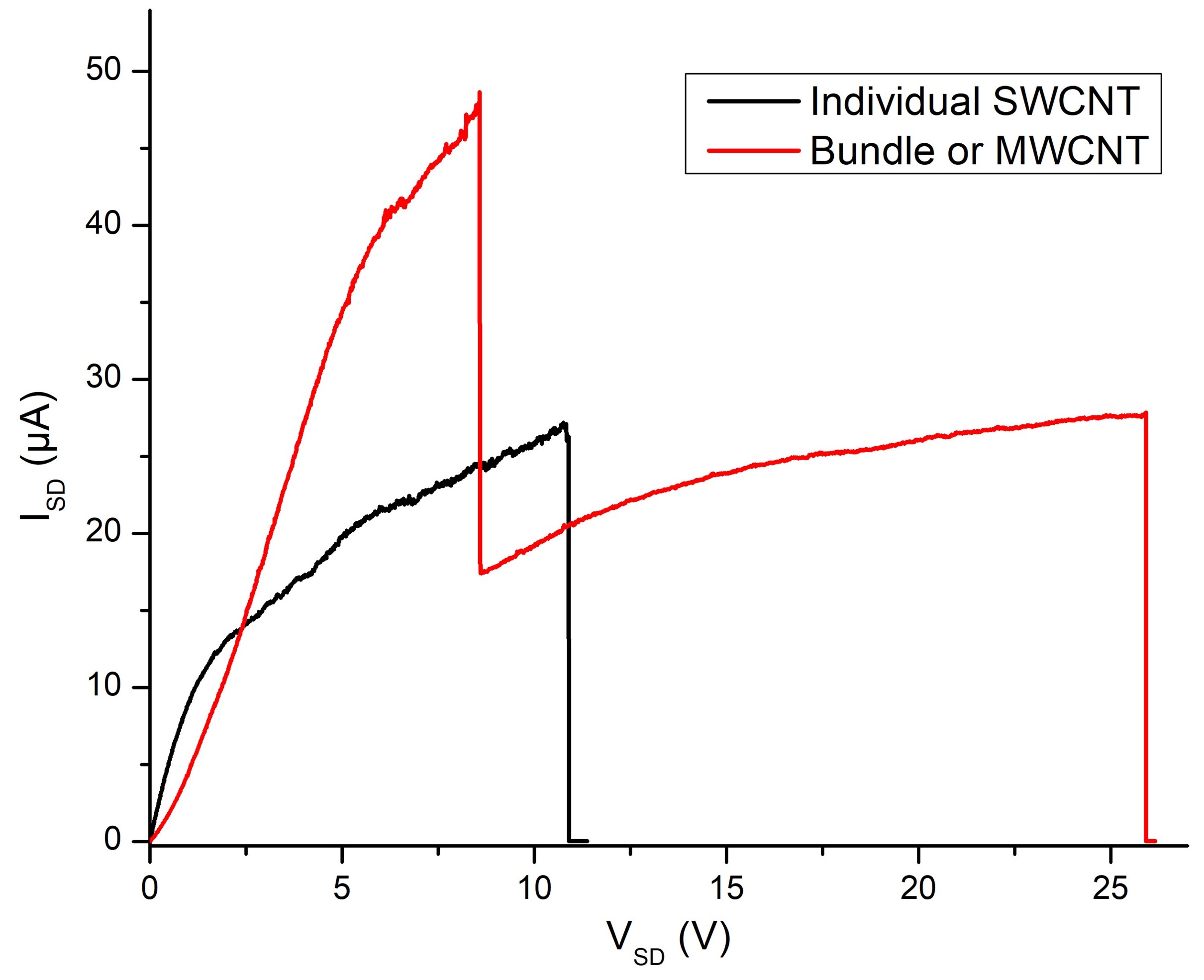}
  \captionsetup{font={small, it}}
  \caption{CNT sides opening and CNT selection. Individual SWCNTs with breakdown current below 30 $\mu A$ and presenting one cut (black color). Bundle or MWCNT with higher breakdown current presenting more cuts (red color).}
  \label{fig:FigureS1}
\end{figure}

\addcontentsline{toc}{section}{1. CNT sides opening and selection}
CVD-grown CNTs are ends closed with fullerene-like caps. In order to fill the CNT, we first need to open its ends. One of the most widely used methods to open CNT ends is plasma etching. However, this method is a rough and dirty process as it requires use of photolithography.\\
In this work, we open CNT sides by applying voltage between selected electrodes aiming to cut CNT at a specific position. Flowing current leads to Joule heating of the CNT up to cutting, observed as a sudden breakdown in the current-voltage curve (Figure \ref{fig:FigureS1}). When we open several CNT devices, we observe that there are two groups of CNTs. CNTs with a breakdown current below 30 \textmu A and presenting one jump are individual SWCNTs, whereas CNTs with breakdown current which is higher (and typically a multiple of the individual case) and present multiple jumps can be bundle or MWCNT.

\section{2. The impact of different humidity levels on the SWCNT transfer characteristic}
\addcontentsline{toc}{section}{2. The impact of different humidity levels on the SWCNT transfer characteristic}

\begin{figure}[htbp]
\centering
  \includegraphics[width=0.5\textwidth]{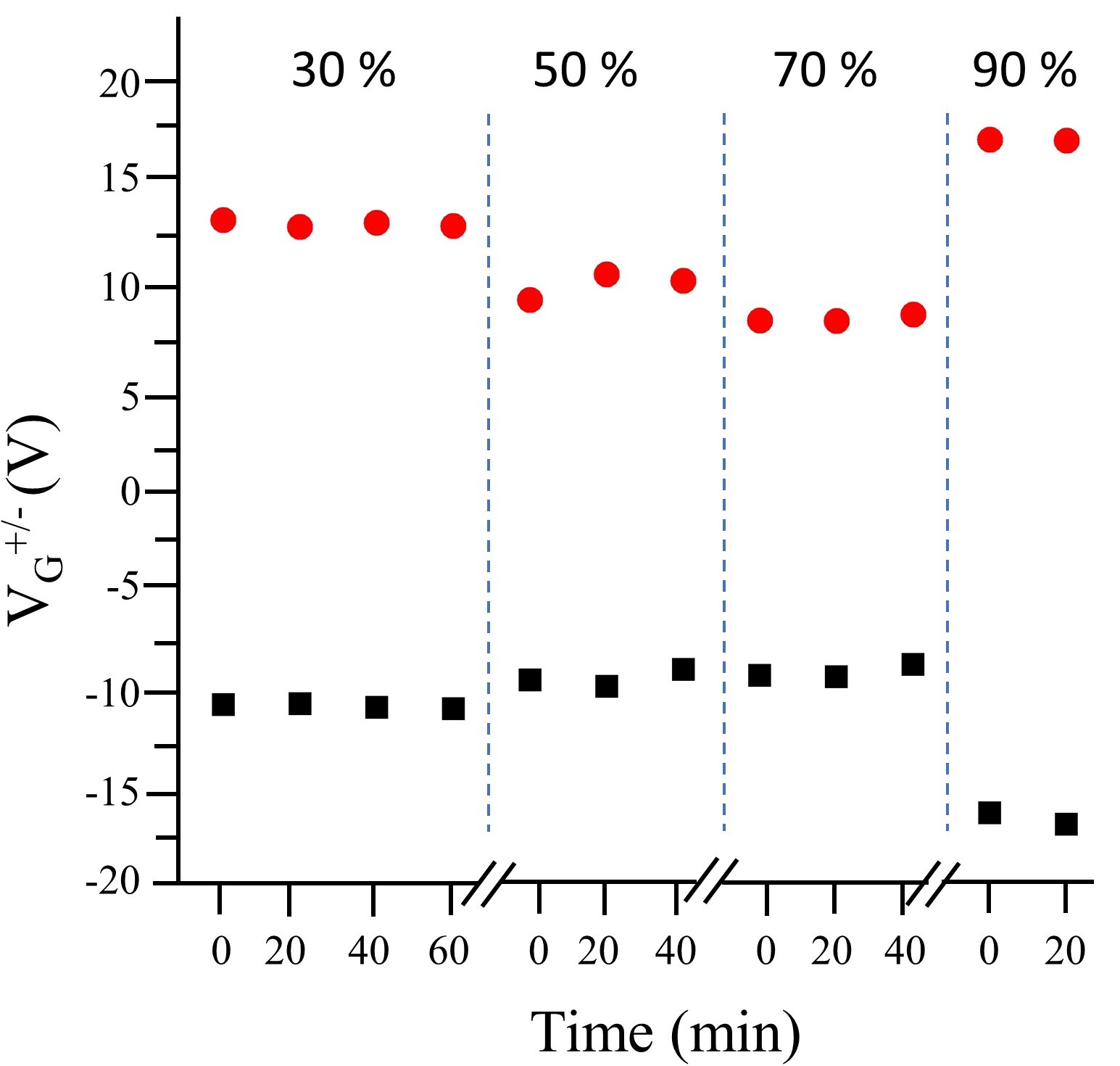}
  \captionsetup{font={small, it}}
  \caption{Impact of humidity on a closed individual SWCNT measured over time for each level of humidity}
  \label{fig:FigureS2}
\end{figure}

Closed individual SWCNT is exposed to different humidity levels from 30\% up to approximately 90\% relative humidity.\\
Figure \ref{fig:FigureS2} reports the change of the gate voltage neutrality points upon device exposure to different humidity levels and recorded over time. We do not observe any significant change at humidity levels up to 70\%. However, at higher humidity, i.e. 90\%,  gate voltage neutrality points shifted towards the extremities of the measurement window. The observed impact for intermediate humidity levels shows that water adsorption outside the CNT already happens at humidity as low as 30\%. However, a higher humidity level is in line with exposure to liquid water, leading to a shift of the neutrality point toward the end of the measurement window, which can be assumed to be caused by water condensation in the tube surrounding.

\section{3. Closed individual SWCNT exposed to different environments in repeated cycles}
\addcontentsline{toc}{section}{3. Closed individual SWCNT exposed to different environments in repeated cycles}

\begin{figure}[htbp]
\centering
  \includegraphics[width=0.5\textwidth]{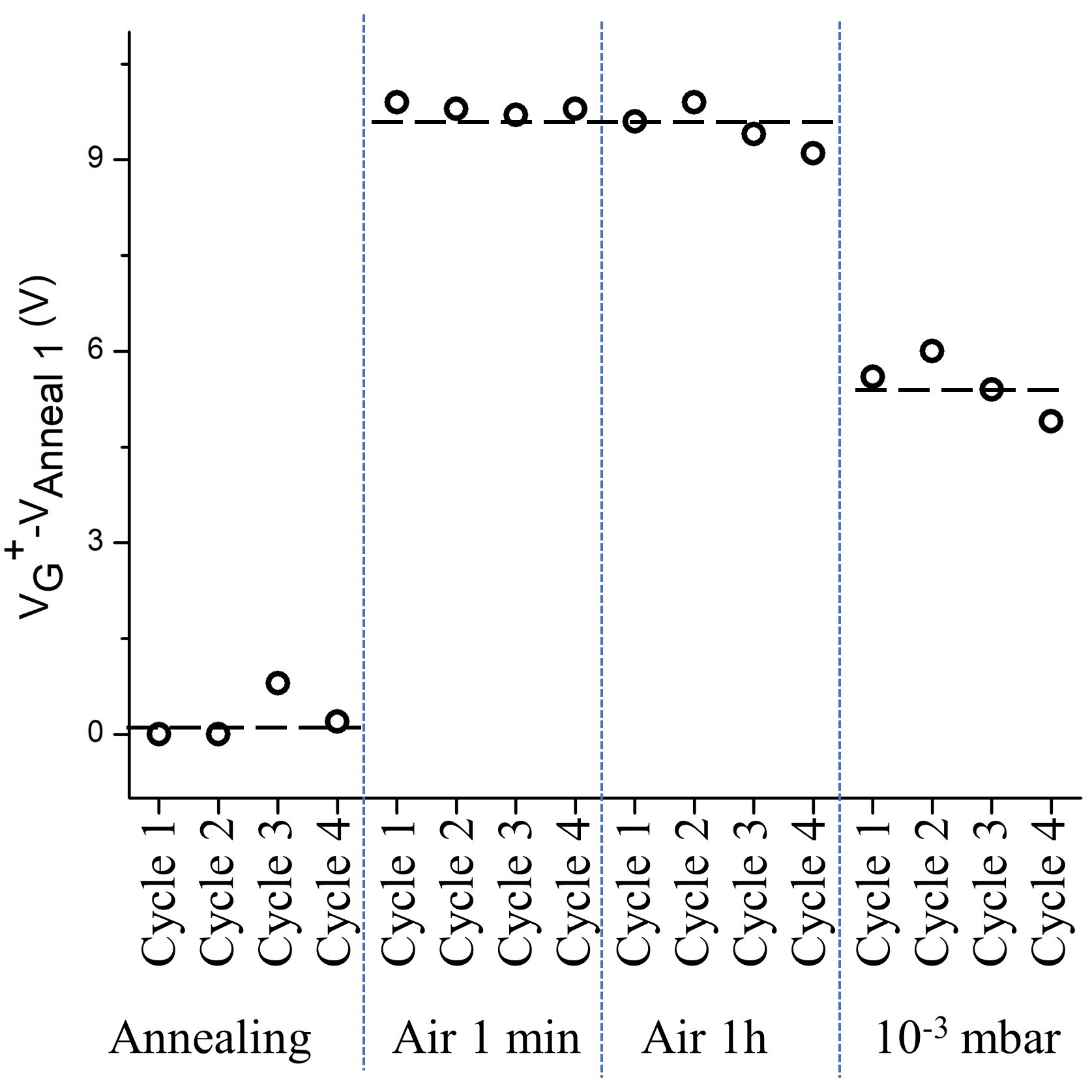}
  \captionsetup{font={small, it}}
  \caption{Evolution of the neutrality point when a closed individual SWCNT is exposed to different environments in repeated cycles.}
  \label{fig:FigureS4}
\end{figure}

In Figure \ref{fig:FigureS4}, the change of the neutrality point upon several cycles of environmental exposure is presented. Each cycle begins with  annealing under vacuum , i.e., in a water-free state, followed by exposure to ambient air for less than 1 minute, then for a long period of around 1 hour, and the cycle ends with placing under vacuum. We conclude that the measurements are reproducible from one cycle of exposure to another. Therefore, the reported effects are not sample history dependent.

\section{4. Exposing the CNTFET to He, and comparison with air}
\addcontentsline{toc}{section}{4. Exposing the CNTFET to He, and comparison with air}

\begin{figure}[htbp]
\centering
  \includegraphics[width=0.5\textwidth]{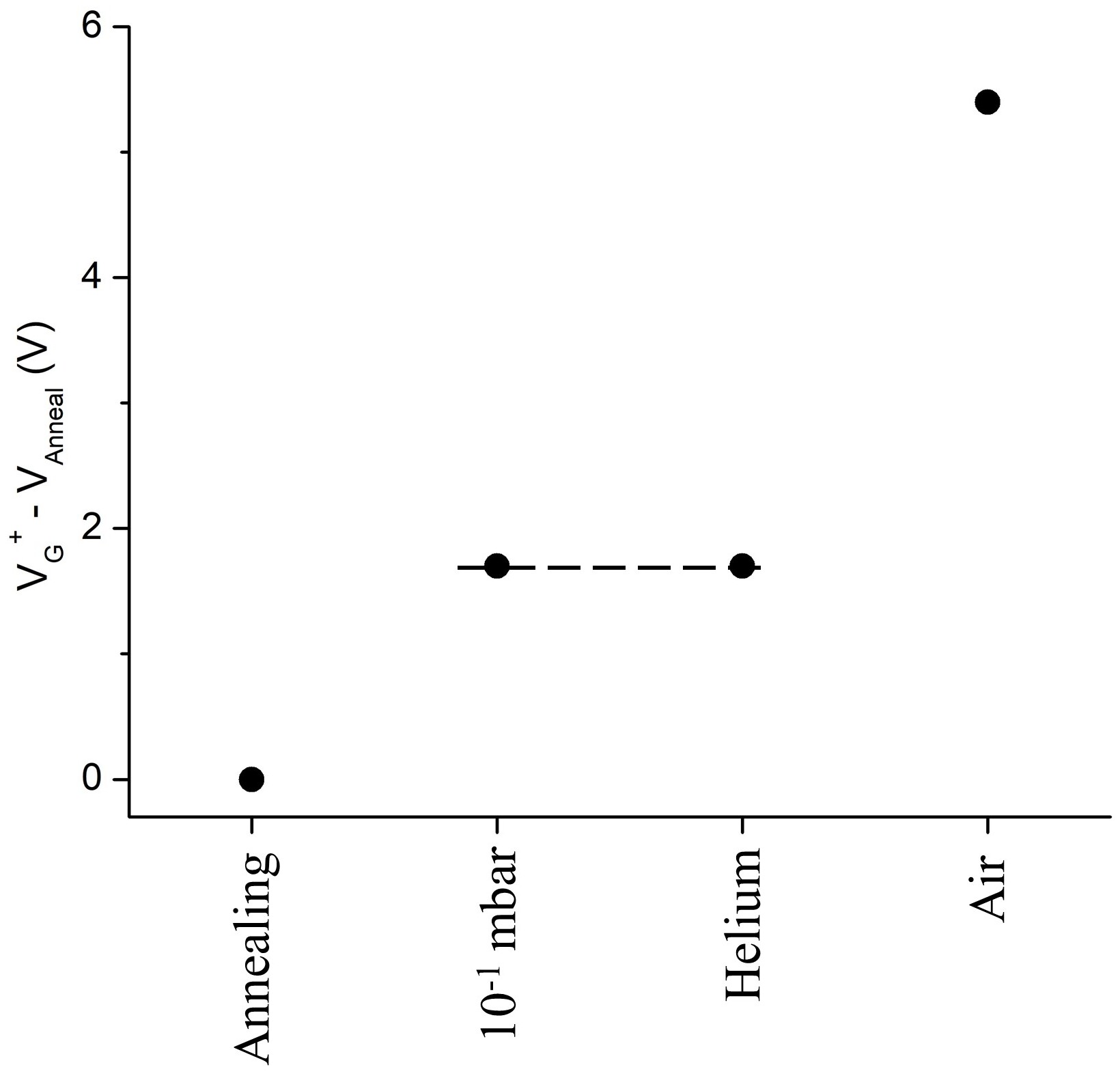}
  \captionsetup{font={small, it}}
  \caption{Closed individual SWCNT exposed to Helium and air after current annealing under vacuum}
  \label{fig:FigureS3}
\end{figure}

We investigated the effect of exposure to air or He. To this aim, a closed individual SWCNT is current annealed under vacuum, then exposed successively to He and air. After annealing under $10^{-3}$ mbar the pump is stopped, effectively leading to a reduction of vacuum to $10^{-1}$ mbar. At this step, we observe a slight shift of the gate voltage neutrality point. Then, we inject He into the chamber. No change in the transfer characteristic is observed (Figure \ref{fig:FigureS3}). Finally, when the device is exposed to air a significant change of the neutrality point V$\textsubscript{G}\textsuperscript{+}$ is measured. Thus it can be concluded that the observed shift upon exposure to air is related to the oxygen or water adsorption on CNT.

\section{5. Liquid water exposure followed by N\textsubscript{2} evaporation}
\addcontentsline{toc}{section}{5. Liquid water exposure followed by N\textsubscript{2} evaporation}

\begin{figure}[htbp]
\centering
  \includegraphics[width=0.5\textwidth]{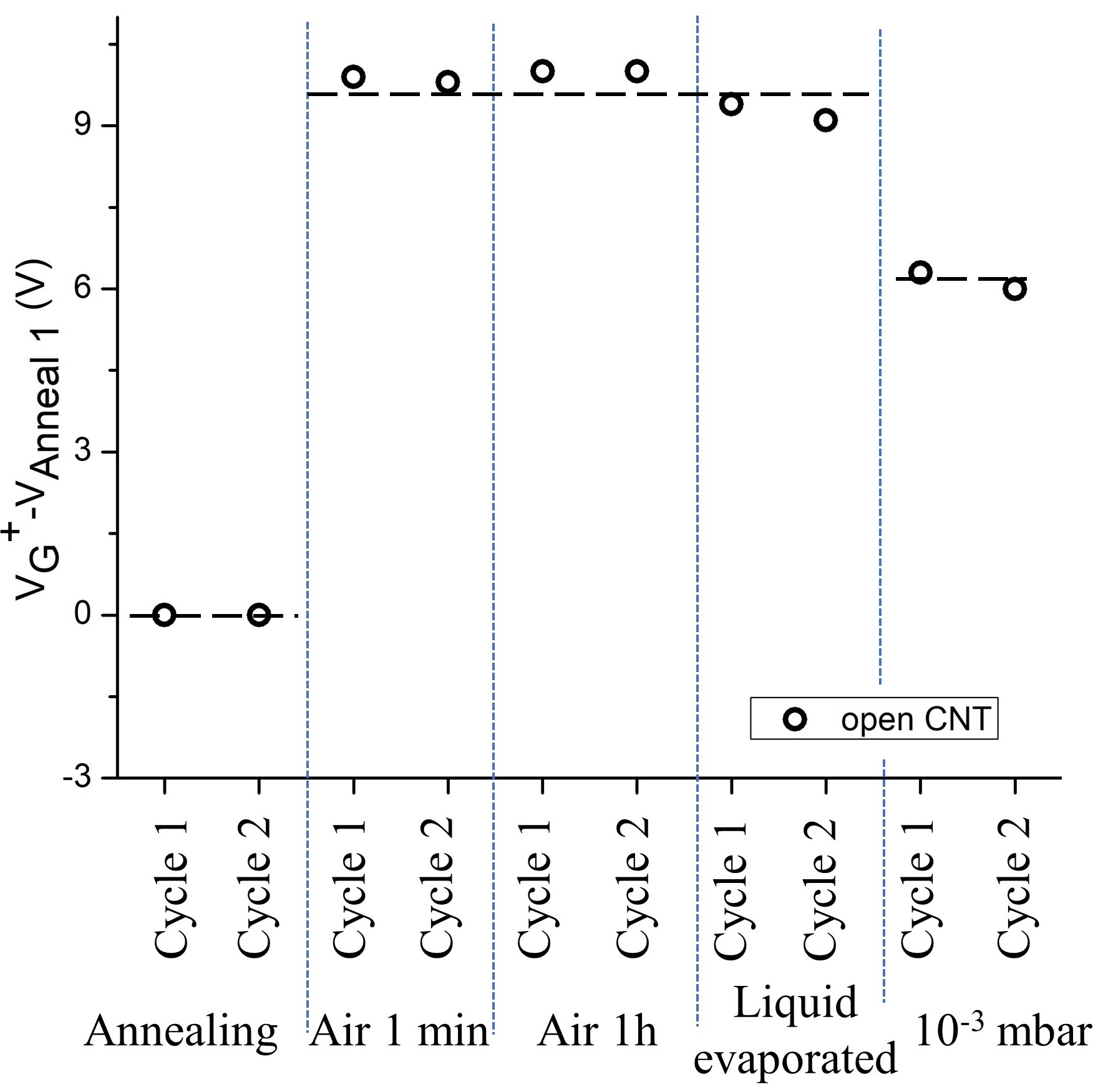}
  \captionsetup{font={small, it}}
  \caption{An open 5 \textmu m long individual SWCNT treated under different conditions, including after CNT being exposed to liquid water and evaporation.}
  \label{fig:FigureS5}
\end{figure}

In addition to the protocol mentioned in the main text, the device with individual SWCNT was thereafter soaked with liquid water and then dried in air. We observed that when the CNT, either open or closed, is exposed to liquid water, the gate hysteresis disappears. However, once the liquid water is evaporated with N\textsubscript{2} flow, the transfer characteristic recovers its initial shape and level (Figure \ref{fig:FigureS5}). Finally, after drying the device is exposed to $10^{-3}$ mbar vacuum, again a decrease of the V$\textsubscript{G}\textsuperscript{+}$ value is observed as discussed in the main text (Figure 3b).
Thus, exposure to liquid water followed by N\textsubscript{2} evaporation had the same impact as CNT exposure to ambient relative humidity in the range of 40-50\% as reported in the main text.

\section{6. Bundle or MWCNT exposed to different environments}
\addcontentsline{toc}{section}{6. Bundle or MWCNT exposed to different environments}

\begin{figure}[htbp]
\centering
  \includegraphics[width=0.5\textwidth]{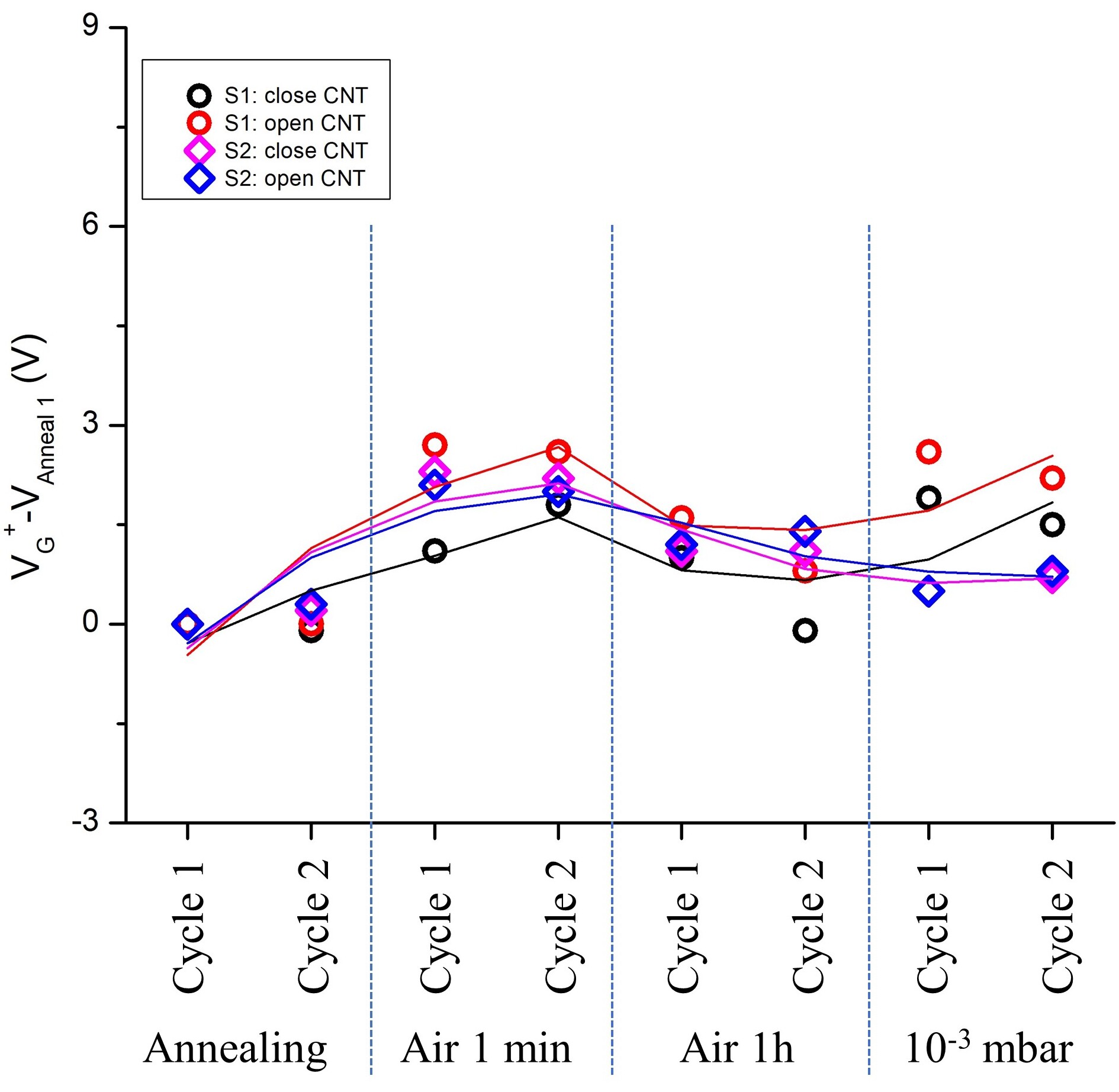}
  \captionsetup{font={small, it}}
  \caption{Comparison of the closed and open cases for two different CNTFETs with the same apparent metallicity made of either bundle or MWCNT. Open symbols are the experimental data obtained on various samples. Lines are just to guide the eyes.}
  \label{fig:FigureS6}
\end{figure}

Devices made of either bundle or MWCNT are identified from the breakdown currents upon opening (see Figure \ref{fig:FigureS1}). Figure \ref{fig:FigureS6} reports the change in the V$\textsubscript{G}\textsuperscript{+}$  values of two devices made of bundle or MWCNT, as a function of the same environmental conditions as for devices made of individual SWCNT (Figure 3a). In these cases, whatever the exact nature of the CNTs, we do not observe any significant change in the neutrality point of the transfer characteristic curves.\\
The observed difference between individual SWCNT and the other group of CNTs when exposed to different environments can be explained in several ways. It is well known that in DWCNT or MWCNT, the measured electronic properties are those dominated from the outermost wall\cite{Liu_2009_DWCNT}. Therefore, we can assume that in the case of a MWCNT the conductivity of the external wall is not influenced by the presence of water confined inside the innermost wall. The underlying hypothesis is that water and carbon interaction is either short-distance or screened by the innertubes.\\

In the case of bundles, it is more difficult to distinguish the impact of water owing to the diversity of tubes (i.e. chirality) and the potential mixture of SWCNTs and MWCNTs with which they are made, is likely to level the response of the FET. Furthermore, CNTs arrangement in the bundle can create different water adsorption sites, e.g. between CNTs, with their own adsorption and desorption energies. It is also possible that some of the CNTs are not in contact with metallic electrodes so the impact of the water can not be observed. Therefore it can be concluded that the impact of water on the electrical response of our CNTFET can only be assessed and understood (or properly analyzed) if it is made of an individual SWCNT.

\section{7. Simulated capacitance of a CNTFET soaked in different media}
\addcontentsline{toc}{section}{7. Simulated capacitance of a CNTFET soaked in different media}

The closed individual SWCNT is soaked in liquids with different dielectric constants. The applied electric field on the SWCNT is estimated by simulating the value of capacitance formed in the case of different liquids (Table \ref{tbl:capacitance}).

\begin{table}[ht]
  \centering
  \captionsetup{font={small, it}}
  \caption{The capacitance formed on the CNT in the case of different liquids}
  \label{tbl:capacitance}
  \begin{tabular}{|c|c|c|}
    \hline
    Medium & Dielectric constant &Maxwell capacitance [F] \\
    \hline
    Air & 1 & $7.49*10^{-17}$ \\
    \hline
    Silicon oil & 2.96 & $1.09*10^{-16}$ \\
    \hline
    Anisole & 4.1 & $1.27*10^{-16}$  \\
    \hline
    Ethyl acetate & 6 & $1.55*10^{-16}$ \\
    \hline
    Water & 80 & $6.4*10^{-16}$\\
    \hline
  \end{tabular}
\end{table}

\bibliography{My_biblio}